**Imaging Spin Dynamics in Monolayer $WS_2$ by Time-Resolved Kerr Rotation Microscopy**

*Elizabeth J. Bushong,*[1] *Michael J. Newburger,*[1] *Yunqiu (Kelly) Luo,*[1] *Kathleen M. McCreary,*[2] *Simranjeet Singh,*[1] *Iwan B. Martin*[1]*, Edward J. Cichewicz Jr.*[1] *Berend T. Jonker,*[2] *Roland K. Kawakami*[1]*

[1]*Department of Physics, The Ohio State University*, Columbus OH 43210, USA
[2]*Naval Research Laboratory*, Washington DC 20375, USA

*e-mail: kawakami.15@osu.edu

Abstract

Monolayer transition metal dichalcogenides (TMD) have immense potential for future spintronic and valleytronic applications due to their two-dimensional nature and long spin/valley lifetimes. We investigate the origin of these long-lived states in n-type $WS_2$ using time-resolved Kerr rotation microscopy and photoluminescence microscopy with ~1 μm spatial resolution. Comparing the spatial dependence of the Kerr rotation signal and the photoluminescence reveals a correlation with neutral exciton emission, which is likely due to the transfer of angular momentum to resident conduction electrons with long spin/valley lifetimes. In addition, we observe an unexpected anticorrelation between the Kerr rotation and trion emission, which provides evidence for the presence of long-lived spin/valley-polarized dark trions. We also find that the spin/valley polarization in $WS_2$ is robust to magnetic fields up to 700 mT, indicative of spins and valleys that are stabilized with strong spin-orbit fields.

I. Introduction

Transition metal dichalcogenide (TMD) monolayers are of great interest due to their two-dimensional (2D) nature combined with their unique band structure. While bulk TMDs have an indirect band gap, the band structure transitions to a direct gap at the K point as the van der Waals coupled layers are isolated down to a single layer[1,2]. In addition, strong spin-orbit coupling due to the transition metal atom creates large spin splitting in the band structure[3-5], as shown in Figure 1a, leading to spin/valley optical selection rules and long spin and valley lifetimes[6-15]. In the tungsten compounds, $WX_2$, optical excitation couples the upper level of the valence band to the upper level of the conduction band. This allows for the formation of "dark" excitons and trions (i.e. charged excitons), a lower energy state that cannot be optically excited or radiatively recombined, which should have even longer lifetimes[16-18]. These properties make TMDs extremely attractive for the field of spintronics, where 2D materials with long lifetimes are necessary for the implementation of nanoscale spintronic devices.

Utilization of these properties, however, demands a fundamental understanding of spin and valley behavior at the microscopic level. Because the helicity of light couples to the spin and valley polarizations via optical selection rules[6,10], optical probes are ideal for investigating the spin and valley dynamics in monolayer TMDs. One can selectively populate either the K or K' valley using positive or negative helicity light, as illustrated in Figure 1a. When these valley populations decay, they emit light of the corresponding helicity, so that the photoluminescence (PL) circular polarization directly reflects the valley polarization, allowing one to follow their temporal evolution. Studies focused on the spatial dependence of optical properties in these materials found that the PL, Raman, and Kerr rotation are highly dependent on position[19-21]. In addition, early studies utilizing polarization-resolved PL spectroscopy found high retention of circular polarization in $MoS_2$ monolayers (up to 99%), suggesting valley lifetimes exceeding 1 ns[7-9,12]. More recent time resolved Kerr rotation[22] (TRKR) measurements observed much longer spin/valley lifetimes (3-80 ns) in n-type monolayer $MoS_2$ and $WS_2$[23,24] and p-type $WSe_2$[25-28] The origin of these long lifetimes is not well understood, however, with some studies attributing the long-lived spin/valley-polarization to resident electrons and holes[23-27], while others ascribe the long lifetimes to spin/valley-polarized dark trions[28].

In this Letter, we utilize time-resolved Kerr rotation microscopy in conjunction with photoluminescence microscopy to elucidate the origin of the long spin and valley lifetimes in n-type monolayer $WS_2$. Notably, we identify a signature that substantiates the presence of long-lived spin/valley-polarized dark trions in addition to the spin/valley polarization of resident conduction electrons. Using TRKR microscopy with high spatio-temporal resolution (<1 μm, 150 fs), we observe complex spatial dependences of spin and valley density varying on the micron length scale, with lifetimes exceeding 5 ns. To understand the origin of the long-lived TRKR signal, which is sensitive to the net spin and valley polarization coming from resident carriers, dark and bright neutral excitons, and dark and bright trions, it is useful to contrast with PL, which is only sensitive to the recombination of bright excitons and trions. Comparing spatial maps of PL microscopy with TRKR microscopy reveals that the neutral exciton PL intensity and the TRKR signal exhibit a correlation, which we attribute to the resident conduction electrons. In addition, we discover an unexpected anticorrelation between trion PL and the TRKR signal, which is explained by the formation of spin/valley-polarized dark trions. This provides new insights on the origin of the long-lived signal and its relationship to both resident electrons and dark trions. We also find that the spin lifetime in $WS_2$ is robust against external magnetic fields due to the stabilization provided by strong spin-orbit coupling. However, a small component of the spin signal is discovered to precess, indicating the existence of an additional spin population whose behavior is not dominated by the spin-orbit field. These results demonstrate high resolution imaging of spin dynamics as a powerful tool for investigating spin-dependent physics in 2D materials.

## II.  CVD Grown $WS_2$

The monolayer flakes of $WS_2$ are grown by chemical vapor deposition (CVD) on $SiO_2$/Si substrates[29]. As shown in Figure 1b, a given $WS_2$ sample has small isolated triangles, typically 10-40 microns in size, which are believed to be single-crystalline. Using PL spectroscopy, we verify the monolayer nature of the $WS_2$ by confirming the lack of indirect gap PL emission at longer wavelengths (~775-900 nm), which is present for bilayer and thicker samples[1,2]. Representative low temperature (8 K) photoluminescence spectra from five different $WS_2$ flakes are shown in Figure 1c, labeled Flake 1-5, respectively. These spectra were chosen in order to show all luminescence features present in the sample, aka maximum number of peaks, although the intensity of each feature varies depending on position. The neutral exciton ($X^0$), trion ($X^-$), and localized defect emission ($L_D$) are labeled in each spectrum. Transport measurements (supplementary material, section I) on similarly prepared $WS_2$ monolayers in a field effect transistor (FET) geometry indicate that the as-grown material is lightly n-doped. This is further substantiated through a comparison of the PL and reflectivity spectra, which exhibits a small Stokes shift (<2 nm) of the exciton peak, consistent with low doping levels[30] (see supplementary material section I for details).

## III.  Time Resolved Kerr Rotation Microscopy

Spin and valley dynamics in monolayer $WS_2$ are investigated using TRKR microscopy, depicted in Figure 1d. The sample is held at 8 K in a low vibration, closed-cycle optical cryostat. An optical parametric oscillator (~150 fs, 76 MHz) is tuned to the energy with the maximum Kerr rotation signal, typically at the energy with the strongest emission peak (flake 1, $\lambda_{Kerr}$ = 608 nm; flake 2, $\lambda_{Kerr}$ = 610 nm; flake 3, $\lambda_{Kerr}$ = 610 nm; flake 4, $\lambda_{Kerr}$ = 606 nm; flake 5, $\lambda_{Kerr}$ = 625 nm). A wavelength dependence of the Kerr rotation in comparison with PL and absorption can be found in Supplementary Section II. Ultrafast pulses from the optical parametric oscillator are split into pump and probe pulses, each of which is focused onto the sample with ~1 μm spot size. The circularly-polarized pump pulse creates valley-polarized excitons, each consisting of a spin polarized electron and hole, as shown in the band diagram (Figure 1a). The time-delayed, linearly-polarized probe pulse measures the combined spin and valley polarization through the Kerr rotation of its linear polarization axis[23,24].

Figure 1e shows representative Kerr rotation, $\theta_k$, as a function of time delay between the pump and probe pulses. As shown in the inset of Figure 1e, there is an initial rapid exponential decay of 3 ps (curve fit in green), which is attributed to the loss of valley polarization of excitons, consistent with previous TRKR studies[31-33]. However, a substantial Kerr rotation remains beyond the initial decay and persists beyond several nanoseconds. We find that the Kerr rotation (after the initial 3 ps decay) in Figure 1e is well described by a bi-exponential decay with time constants of $\tau_{short}$ = 320 ps and $\tau_{long}$ = 5.4 ns (the curve fit is the solid red line in Figure 1e). Bright excitons and trions could not produce the long-lived signal because they recombine within the first few hundred picoseconds,[34-36] but the signal may be due to the spin and/or valley polarization of resident conduction electrons[23-25] or dark trions[16-18,28]. Such bi-exponential behavior is consistent with other TRKR studies of monolayer TMDs[26,27,37]. In monolayer $MoS_2$, which has small spin orbit splitting of the conduction band (~3 meV or less)[4], previous studies indicate that the TRKR signal is dominated by electron spin polarization[23,24]. In the limit of very large spin-orbit splitting, such as that observed in the valence band of TMD monolayers (150 meV – 450 meV)[3], the spin-valley locking will cause the spin to be accompanied by an equivalent valley polarization[25,27,28]. The conduction band of monolayer $WS_2$ is an intermediate case (splitting of ~30 meV)[4], so the spin polarization could be accompanied by a smaller, but non-negligible component of valley polarization. In fact, if the electron density were smaller than the splitting of the conduction band, the signal could be dominated by valley polarization. Therefore, we refer to the TRKR signal as the "spin/valley density."

## IV. High-Resolution Imaging of Spin and Valley Dynamics

We investigate the spatial dependence of the spin/valley density by obtaining high-resolution maps of the TRKR signal. By scanning the overlapped pump and probe beams at a fixed time delay, a spatial image of the Kerr rotation is obtained. Repeating this at different time delays reveals the microscopic evolution of the spin and valley dynamics. Figure 2 shows a series of Kerr rotation images taken on a triangular island (flake 5) of monolayer $WS_2$ at different time delays. The TRKR excitation and detection are at 625 nm, which corresponds to the brightest peak in the PL spectrum (see Figure 1c). The sequence of images illustrates a complex spatial dependence of the TRKR signal on the $WS_2$ island. Areas with a large signal (colored in yellow/orange/red) are separated by only a few microns from regions with almost no signal (colored blue/black). The spatial distribution is striking, with a central core of low spin/valley density surrounded by regions of higher spin/valley density. It is worthwhile to note that even at 11 ns, there is still a measurable signal. In addition, we investigate the sample-to-sample variation of the spin/valley density pattern by mapping the Kerr rotation of many other

WS$_2$ flakes, shown in Figure 3a. While we observe a variety of spin/valley density spatial dependences, all flakes have large variations in the intensity of the signal over the distance of a few microns.

## V. Comparison of Photoluminescence Microscopy and TRKR Microscopy

To gain further insight into the spatial dependence of the Kerr rotation, we investigate its relationship with the spatial dependence of the individual neutral exciton, trion, and defect peaks. Figure 3 shows the TRKR spatial maps for five different WS$_2$ flakes (Figure 3a) along with the corresponding maps of photoluminescence integrated over the full PL spectrum (Figure 3b), the map of the neutral exciton peak amplitude (Figure 3c), the map of the trion peak amplitude (Figure 3d), and the map of the defect peak amplitude (Figure 3e).

We first analyze the relationship between the TRKR map and the full spectrum integrated PL. Comparing the TRKR maps across the different flakes (Figure 3a), it is apparent that the spatial dependence of spin/valley density varies from sample to sample and presents multiple patterns. Similar spatial patterns are seen in the TRKR (Figure 3a) and full spectrum PL maps (Figure 3b) for each flake, indicating a relationship between spin/valley density and photoluminescence. However, when the relationship between each TRKR map and its corresponding PL map (same column) is more carefully examined, two different behaviors appear. In flakes 1, 2, and 4, areas of high spin/valley density appear to be positively correlated to areas of large PL intensity (Figure 3b). This is reasonable when considering that both TRKR and PL signal strength should benefit from material with better optical quality. In contrast, regions of flakes 3 and 5 exhibit an unexpected anticorrelation between spin/valley density and PL. This anticorrelation is also evident in a particular spot in flake 1 (circled in white), which developed after laser exposure (A discussion of this can be found in supplementary section III). In flake 3, PL increases while TRKR decreases in the center of the flake, but the outer regions show PL and TRKR increasing/decreasing together. In flake 5, the anticorrelation manifests more sharply in certain regions, particularly in a spot at the top right of the flake. Here, individual PL and TRKR lifetime scans (supplementary section IV) reveal that the area of strongest PL exhibits the expected fast decay (<10 ps) associated with the loss of valley polarization of excitons but shows little amplitude in the long-lived spin/valley state. On the contrary, regions with high spin/valley density in the long-lived state reveal much weaker PL. While this variance across flakes between correlation and anti-correlation of TRKR and full spectrum PL does not have a clear explanation at first glance, investigating the spatial dependence of the individual PL peaks leads to a better understanding of the mechanisms that result in long spin/valley lifetimes.

We further investigate the origin of the long-lived states by comparing the spatial dependence of the neutral exciton peak with the TRKR map. As shown in Figure 1c the characteristic PL spectra from these flakes are composed of emission peaks for neutral excitons (green), trions (blue), and defects (purple). We find that all the flakes in this study exhibit resolvable A exciton luminescence in the regime of ~610 nm, although with varying intensities. Comparing the maps of the neutral exciton peak amplitude (Figure 3c) with the TRKR spatial dependence, it is clear that the $X^o$ emission and TRKR signal are strongly correlated in most regions. The exceptions are the center of flake 3 and a spot in flake 1, where the role of trions dominates the TRKR, as discussed later. The correlation with the exciton is particularly striking for flake 5, where the full spectrum PL is dominated by trion emission and does not have the same spatial dependence as the TRKR, yet the exciton PL still clearly correlates with the TRKR. As stronger exciton emission is generally associated with better material quality, the correlation with the neutral exciton observed in all flakes is expected

The spatial dependence of the trion peak amplitude provides the last piece of the puzzle for understanding the anticorrelation. When comparing the trion maps (Figure 3d) with the TRKR maps for flake 2 and the majority of flake 1, the amplitude of the trion peak appears strongly correlated with the TRKR signal. However, a different behavior is observed in regions of flakes 3, 5 and one spot in flake 1 where the trion intensity shows an anticorrelation with the TRKR strength. In flake 5, the trion emission dominates the PL spectra throughout the entire flake (the spectrum shown in Figure 1c is a location with very low trion emission relative to the other areas of the flake, but the trion is still the dominant feature). Notably, the trion map shows that trion emission is strongest in the top right corner of the flake, the same region where the TRKR map shows a sharp decrease in spin/valley density. Similarly, examining the maps of the individual emission peaks show that the anticorrelation behavior observed between the TRKR and the full spectrum PL in both flake 3 and 1 also originates from the trion peak emission, where the opposite trends between TRKR and PL appear with sharpest contrast in the trion PL map.

Throughout the study, the anticorrelation of the TRKR intensity and photoluminescence only happens when the trion peak is present, and only with the trion peak. The neutral exciton emission is correlated with the TRKR signal across all samples, regardless of the existence of a trion peak. This is likely due to the transfer of angular momentum to resident conduction electrons[23,24]. However, in samples where the trion peak is comparable to or larger than the neutral exction peak, an anticorrelation may be observed. The fact that this behavior is only present in samples with relatively large trion emission (in some well-defined regions) suggests that the anticorrelation may occur in more n-type material, where bright trions are created more efficiently upon optical excitation[25,38]. This behavior can therefore be explained by the conversion of spin/valley-polarized bright trions to non-radiative, dark trion states as follows[17,18,28]. There are four possible mechanisms in which this could occur, as demonstrated in Figure 4. Since the material is slightly n-type, the lower conduction band in both valleys has occupied states with spin opposite that of the valence band in the respective valley. To simplify the following discussion, we assume that optical excitation occurs within a single valley, say the K valley, through the absorption of circularly polarized light, leading to an initial (bright) trion state (e.g. Figures 4a or 4d). The first two methods for producing a dark trion, shown in figure 4b and 4c, begin through excitation of a singlet trion, diagrammed in Figure 4a. In this case the trion is entirely in the K valley. If the trion then decays to the lower energy dark trion through momentum scattering to the K' valley (Figure 4b), the trion can no longer radiatively relax, causing a decrease in trion PL. Since the absorption of linearly polarized light will be affected by the composition of both valleys, a Kerr rotation will persist. Alternatively, the singlet trion can decay to a dark state through the transfer of spin angular momentum to resident electrons through a spin flip to the lower conduction band (figure 4c, the conduction band has a strong $d_{z^2}$ character). This trion is also non-radiative, leading to suppression of the PL. However, it would also be accompanied by a long-lived Kerr rotation due to the momentum of the resident conduction electrons, as well as any change in absorption from the altered composition of the valleys.

The second two methods which result in dark trions, shown in figures 4e and 4f, begin from the triplet trion state, diagrammed in figure 4d. The triplet trion occupies both valleys, being composed of an optically excited electron and hole in the K valley, and an electron in the lower conduction band of the K' valley. From the triplet trion, the same processes can occur as with the singlet trion. The optically excited electron can momentum scatter to the other valley, shown in figure 4e, or it can spin flip to the lower conduction band, figure 4f. In both cases the trion becomes dark, resulting in suppressed trion PL while a Kerr rotation signal will remain. While the four scenarios all fit the observed behavior, the singlet trion decaying via intervalley scattering (Figure 4b) is the most energetically favorable[39]. Regardless of the particular scenario, the conversion of bright to dark trions provides a natural explanation for the anticorrelation between the Kerr rotation and the trion emission maps, where efficient (inefficient) conversion results in lower (higher) trion emission and a higher (lower) Kerr rotation.

The existence of spin/valley-polarized dark trions in WS$_2$ can therefore explain the observed anticorrelation.

We also consider the possibility of dark neutral excitons contributing to the TRKR signal. The presence of such species could also be identified through an anticorrelation with the neutral exciton PL peak. However, we have only observed correlation with the neutral exciton PL peak, and we have never observed anticorrelation behavior in samples without trion peaks. While we cannot rule out the possibility that spin/valley-polarized dark neutral excitons are present, we also do not see any positive evidence to support this. The combination of bright to dark trion conversion combined with the transfer of angular momentum to resident conduction electrons discussed earlier account for both the observed anticorrelation and correlation behavior, making the co-existence of both mechanisms probable in the WS$_2$ samples.

## VI. Effect of Spin-Orbit Coupling on Spin/Valley Dynamics

We now turn our attention to the important role that strong spin-orbit coupling has on the spin relaxation and dynamics in monolayer WS$_2$. Specifically, there are predictions that large out-of-plane spin-orbit fields will enhance spin lifetimes through two primary effects[8,10,13-15]. First, the absence of in-plane components to the spin-orbit field will enhance the lifetime of out-of-plane spins. This has been demonstrated previously for GaAs(110) quantum wells, which exhibit an order of magnitude longer spin lifetime than in GaAs(100) quantum wells[40]. Second, the presence of a strong out-of-plane spin-orbit field can stabilize spins against relaxation induced by external magnetic fields. For this, monolayer WS$_2$ is quite different from the GaAs(110) quantum wells because the spin-orbit field has a large non-zero value at the conduction band minima (~30 meV calculated spin splitting[4]) whereas GaAs(110) quantum wells have a spin-orbit field that goes to zero at the conduction band minimum[41]. This spin stabilization has been observed in p-type WSe$_2$ in external magnetic fields up to 300 mT[25,27,28], however the valence band spin splitting in WSe$_2$ is an order of magnitude larger (~450 meV)[3] than the spin-splitting in the conduction band of WS$_2$ (~30 meV)[4], placing WS$_2$ in an intermediate regime. It is therefore important to experimentally investigate the spin/valley dynamics in WS$_2$ as a function of external magnetic field.

The results for TRKR on WS$_2$ as a function of B$_{ext}$ (Figure 5a) indicate that the dynamics are governed by spin/valley stabilization resulting from strong spin-orbit coupling. The data presented was obtained from flake 5 on a spot with high spin/valley density based on a TRKR map. The observed behavior is consistent across multiple flakes and samples, and we do not observe a systematic trend of the TRKR amplitude as a function of magnetic field. The spin/valley lifetimes, obtained by fitting the data in Figure 5a, exhibit very little dependence on $B_{ext}$ (Figure 5b). The lack of large oscillations, such as those seen in typical semiconductors, and the robustness of the spin/valley lifetime to external fields are both characteristic of the spin-orbit stabilization. Comparison of other TMD properties to $WS_2$ provides additional support for the robustness of the spin polarization originating from spin stabilization due to strong spin orbit coupling. While the spin/valley lifetime in $WS_2$ shows little variation up to 700 mT, the spin lifetime of $MoS_2$, which has much weaker spin-splitting in the conduction band (~3 meV in $MoS_2$, ~30 meV in $WS_2$)[4], was shown to collapse in fields as low as 100 mT[23,24]. In addition, the temperature dependence of the TRKR signal in $WS_2$ in comparison with other TMDs supplies additional evidence for spin stabilization. The spin lifetime in $WS_2$ persists to 130 K, beyond which it drops to <200 ps (see supplementary material section VI). In comparison, the spin lifetime of electrons in $MoS_2$ is much less robust, falling to <200 ps at only 40 K[24]. In p-type $WSe_2$, where lifetimes on the order of ~80 ns have been observed[27], spin/valley polarization persists up to room temperature[26]. The behavior for p-type monolayer TMDs should be qualitatively different than for the n-type material because the very large spin splitting of the valence band should promote spin-valley locking and produce long lived spin/valley polarization based on suppression of intervalley scattering.

While the spin lifetime does not depend on an external magnetic field, detailed TRKR delay scans with smaller time steps and more signal averaging (Figure 5c) reveal the presence of a small oscillatory signal (<3% of the total Kerr signal), which is not expected with strong spin-orbit stabilization. The oscillations are due to an additional spin population which contributes to the overall signal. We find that this additional spin population has a g-factor of 1.90 ± 0.04 (by fitting $\nu_L$ vs. $B_{ext}$) and a dephasing rate $1/T_2^*$ that grows linearly with $B_{ext}$ (Figure 5e). These dynamical properties are consistent with a spin population where dephasing is governed by inhomogeneous broadening of the g-factor. Such oscillatory Kerr signals with similar dynamical properties have been observed in monolayer $MoS_2$ and have been attributed to the presence of spins in localized states[23]. Therefore, while most of the Kerr signal comes from non-precessing conduction band spins in the spin-orbit stabilized regime, composed of both resident conduction electrons and trions, a small contribution to the Kerr signal comes from precessing spins in localized states.

In conclusion, we utilize TRKR microscopy to image the dynamics of spin/valley density in monolayer $WS_2$. We discover a complex spatial dependence of spin/valley density in triangular islands that persists beyond 11 ns. Comparing photoluminescence microscopy with TRKR microscopy reveals the role of dark trions in the long lifetimes observed in $WS_2$, particularly in areas with the unexpected anticorrelation. Application of in-plane magnetic fields indicates the presence of two spin populations that contribute to the Kerr rotation signal. A small contribution to the Kerr signal likely comes from precessing spins in localized states, however most of the Kerr signal comes from non-precessing conduction band spins that are stabilized by large spin-orbit coupling against spin relaxation. The successful demonstration of high resolution TRKR microscopy on monolayer TMD enables numerous studies of spin and valley dynamics in TMD films and heterostructures that are crucial for developing spintronic applications.

**Acknowledgements**

The work at Ohio State was primarily supported by NSF (DMR-1310661) and received partial support from NSF MRSEC (DMR-1420451) and C-SPIN STARnet, a Semiconductor Research Corporation program sponsored by MARCO and DARPA. The work at NRL was supported by core programs at NRL and the NRL Nanoscience Institute and received partial support from AFOSR (F4GGA24233G001).

**Author Contributions**

E.J.B., M.J.N., and Y.K.L. performed the optical measurements. K.M.M. performed the CVD synthesis of samples. S.S. prepared samples for measurement. E.J.B., M.J.N., Y.K.L., K.M.M., S.S., B.T.J., and R.K.K. discussed the results and contributed to the manuscript.

**Competing financial interests:** The authors declare no competing financial interests.

**Methods**

**Time-resolved Kerr rotation microscopy**
Time-resolved Kerr rotation microscopy measurements are performed using ~150 fs laser pulses from an optical parameteric oscillator (Coherent OPO) pumped by a Ti:sapphire laser (Coherent Mira) with repetition rate of 76 MHz. The pump beam is helicity-modulated at 50 kHz using a photoelastic modulator (Hinds) and the probe beam is chopped at 493 Hz and linearly polarized with a $5 \times 10^5$ extinction ratio. A Soleil-Babinet compensator is placed in the pump line after the photoelastic modulator to ensure that the polarization is circular at the sample. The time delay between the pump and probe pulses is adjusted using a mechanical delay line. The overlapped beams are tightly focused into ~1 μm spots on the sample using a 100x Mitutoyo objective with 13 mm working distance. Typical pump and probe powers are 100 μW and 100 μW, respectively. We verified that measurements with a 10:1 pump/probe ratio exhibit the same characteristics, however the stronger probe power was employed for better signal-to-noise. The sample is mounted on an XYZ piezo stage in high vacuum at 6 K in a closed cycle Montana Instruments Cryostation with an external electromagnet which can apply up to 720 mT. The rotation of linear polarization of the reflected probe pulse is detected using a photodiode bridge, and the signal is amplified using a voltage pre-amp (Stanford Research 560). Heterodyne detection using two lock-in amplifiers (Signal Recovery 7270) is used for noise reduction and cancellation of the pump beam. For the temperature dependent measurement, the wavelength is tuned to maximize the Kerr rotation at each temperature.

**Photoluminescence microscopy**

The photoluminescence measurements are performed using a 532 nm linearly polarized diode laser excitation source, which is focused down on the sample to ~1 μm using a 100x Mitutoyo objective with 13 mm working distance. The power of the excitation beam is typically 100 μW. In the detection path, the excitation source is blocked using a notch filter, and the photoluminescence spectrum is detected using a 0.55 m spectrometer (Horiba iHR550) and thermoelectrically-cooled, back-thinned CCD camera (Horiba Synapse).

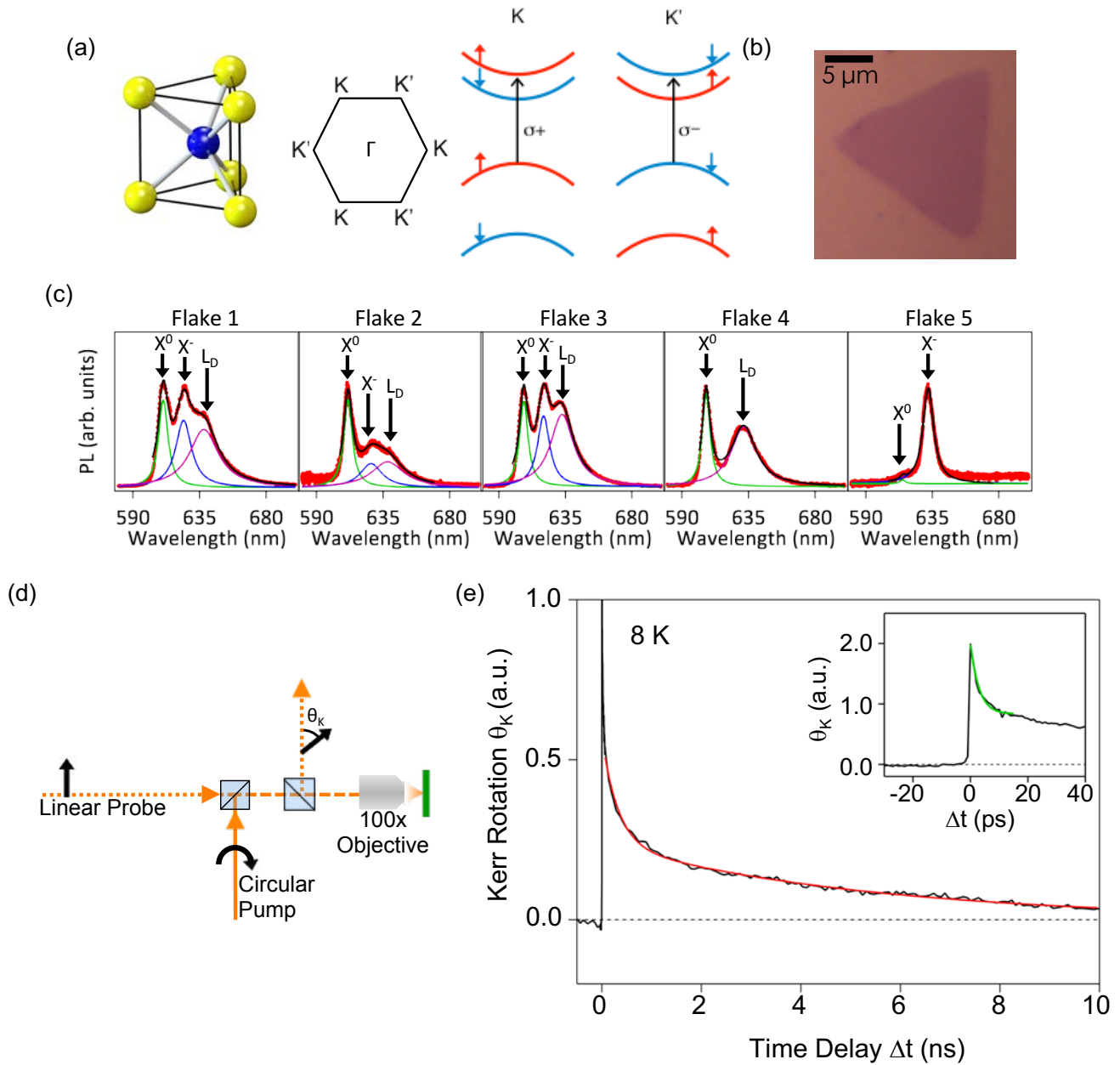

**Figure 1. Time resolved Kerr rotation on high quality CVD WS$_2$ monolayers. a,** (left) The atomic structure of WS$_2$; tungsten is the central blue atom, and the sulfur atoms are yellow. (right) The schematic band structure of monolayer WS$_2$ at the K and -K points. The spin-valley coupling allows one to selectively excite spins in either the K or -K valley. **b,** Optical micrograph of one of the triangular islands used in this study. **c,** Photoluminescence spectroscopy of five monolayer WS$_2$ flakes measured at 8 K are shown in red. In each spectra, the neutral exciton, X$^0$, the trion, X$^-$, and the defect, L$_D$, are labeled. The spectra are fit in the black curve, with the individual peak fits shown in green (neutral exciton), blue (trion), and purple (defect). **d,** Diagram of the TRKR microscopy set-up. **e,** Representative Kerr rotation as a function of pump-probe time delay for monolayer WS$_2$ (flake 5) at 8 K and zero magnetic field. The excitation and detection wavelength was 625 nm. The red curve is a bi-exponential fit yielding time constants of 320 ps and 5.4 ns. Inset: Kerr rotation at short time delays. An exponential fit to the fast decay (green curve) yields a time constant of 3.0 ps.

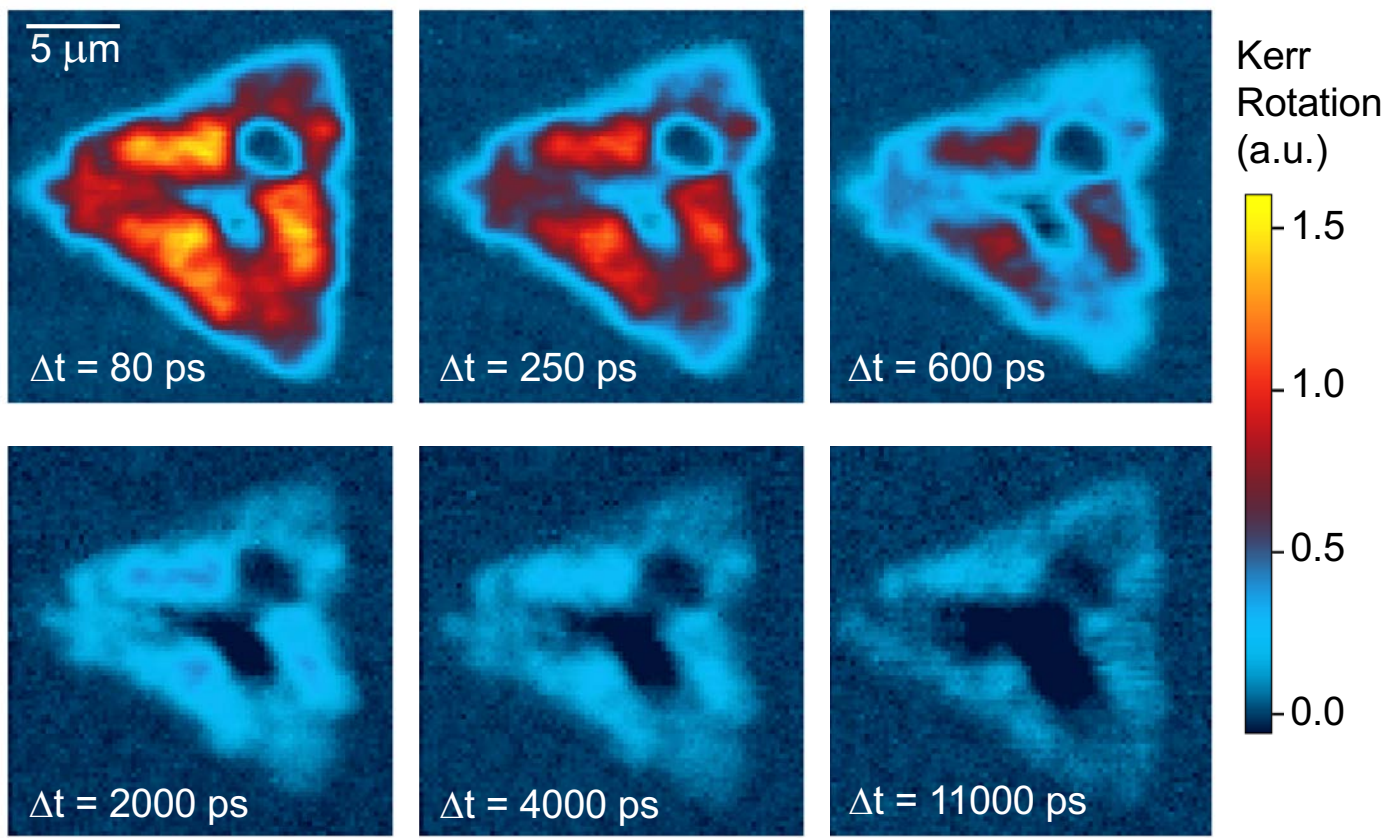

**Figure 2. Spatially-resolved images of time resolved Kerr rotation.** Scanning a WS$_2$ island beneath the overlapped pump and probe beams at a fixed time delay produces a high-resolution spatial map of spin density. The series of images are snapshots of the spin density at time delays of 80, 250, 600, 2000, 4000, and 11000 ps. The images reveal a complex spatial dependence and time evolution of the spins, with regions of high and low spin density in close proximity.

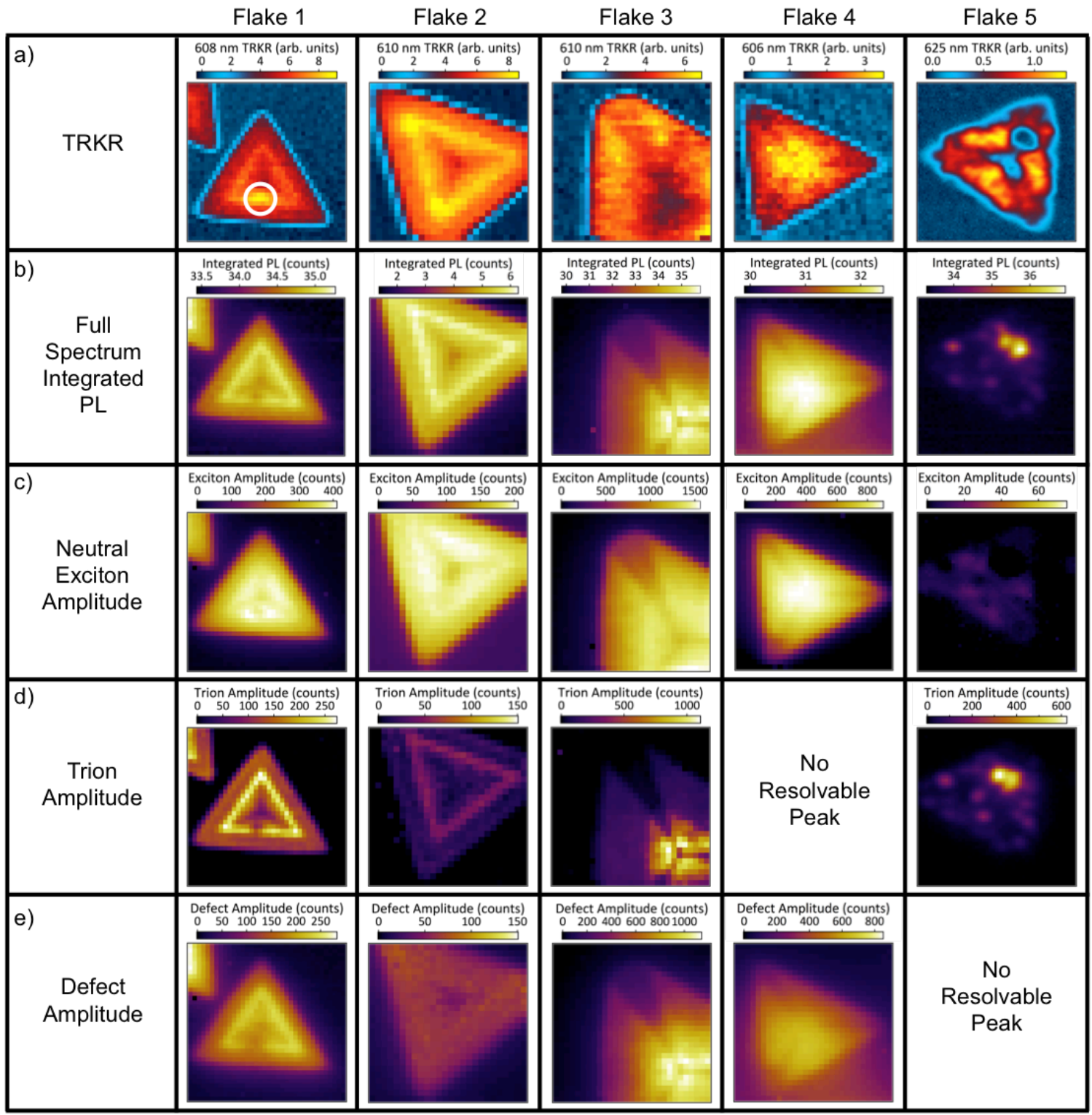

**Figure 3. Kerr rotation and spectrally resolved photoluminescence of WS$_2$ flakes. a,** TRKR maps for each of the five different flakes used in this study are shown in row a. Each map is at a pump-probe time delay of 100 ps. **b,** Maps of the full spectrum integrated PL for each of the five flakes are shown in row b. At each point on the flake, the PL spectra was integrated over the entire emission range. **c,** Maps of the PL only integrated over the exciton range are shown in row c for each flake. When compared with the TRKR maps, it is clear that the exciton PL correlated with the TRKR signal. **d,** Maps of the PL integrated only over the trion range are shown in row d. All data was taken at 8 K.

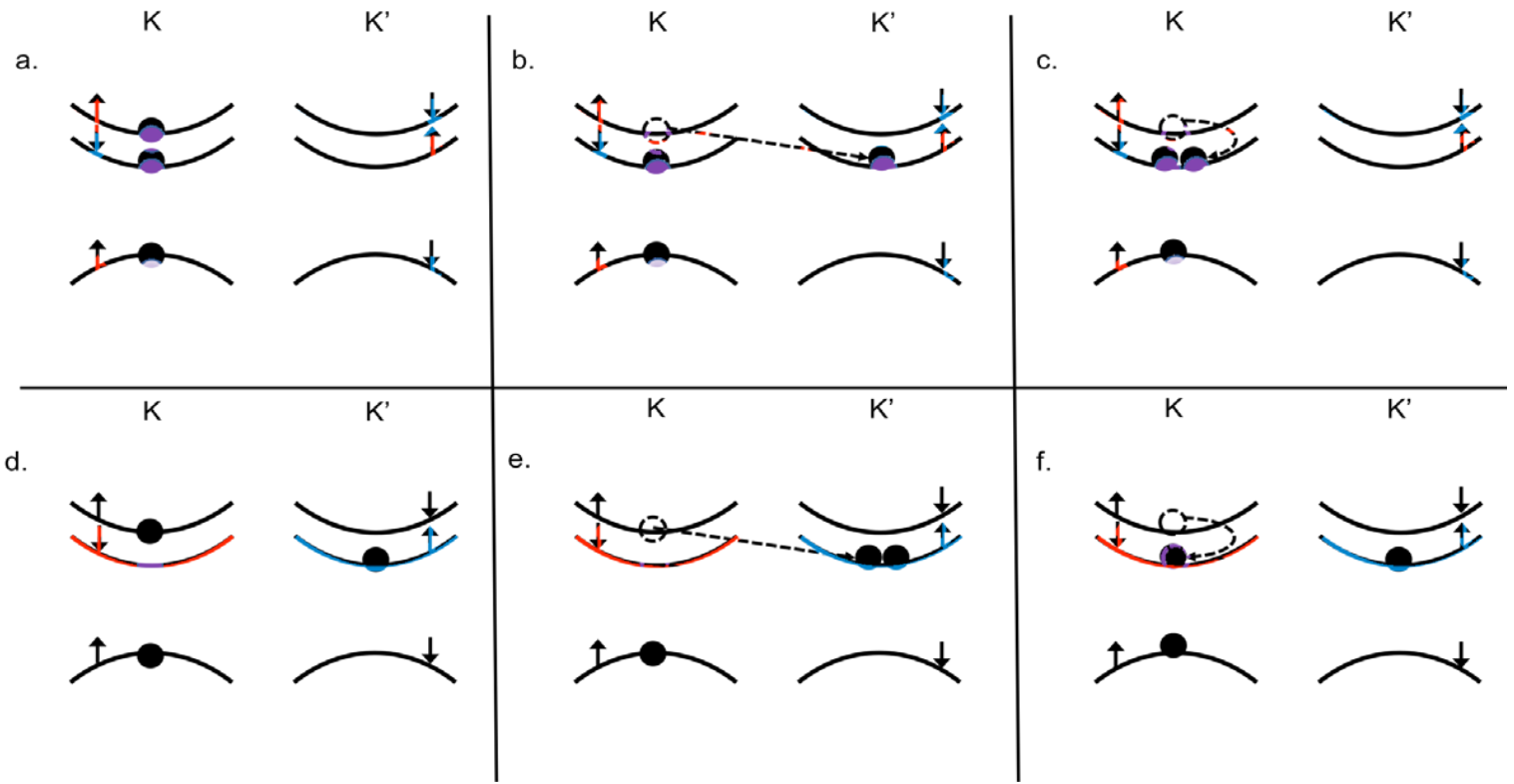

**Figure 4. Bright and dark trion states in $WS_2$. a,** A singlet trion optically excited in the K valley. The electron in the lower valence band is due to n-type doping of the material. **b,** The singlet trion can become a dark trion by momentum scattering of the spin up electron in the K valley to the spin up band of the K' valley. **c,** Another possibility for the singlet trion to convert to a dark trion is though a spin flip to the lower conduction band in the K valley. **d,** A triplet trion. The electron and hole are optically excited in the K valley, and the electron in the K' valley is due to n-type doping. **e,** The triplet trion can convert to a dark trion through momentum scattering of the electron in the K valley to the lower conduction band of the K' valley. **f,** Another possibility for the triplet trion to convert to a dark trion is through a spin flip from the upper conduction band of the K valley to the lower conduction band of the K valley.

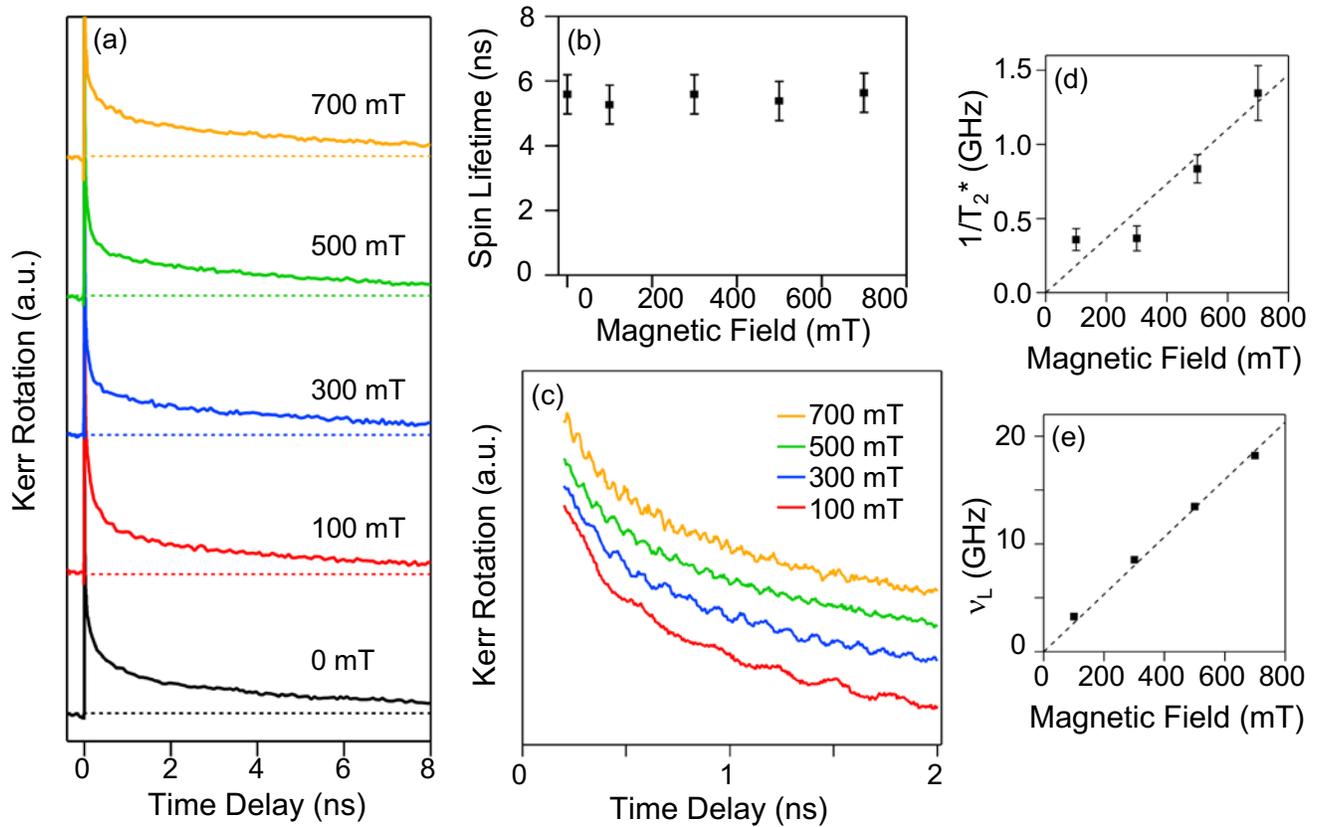

**Figure 5. Spin-orbit stabilized spins in WS$_2$. a,** Normalized Kerr rotation as a function of time delay for different B$_{ext}$. The non-precessing decay curves are characteristic of a spin-orbit-stabilized system. **b,** Spin lifetime as a function of B$_{ext}$, obtained by fitting the TRKR data in (a). The spin lifetime is robust to an external magnetic field up to 700 mT. **c,** Detailed time delay scans of Kerr rotation for different B$_{ext}$. Curves are zoomed in and offset to show a small oscillatory component of the overall TRKR signal (<3%). **d-e,** Larmor precession frequency and dephasing rate, respectively, of the oscillatory component. Both exhibit a linear dependence on B$_{ext}$, which indicate spin dynamics like those of materials without spin-stabilization due to strong spin-orbit coupling.

Supplementary Online Material

# Imaging Spin Dynamics in Monolayer WS$_2$ by Time-Resolved Kerr Rotation Microscopy

*Elizabeth J. Bushong,[1] Michael J. Newburger,[1] Yunqiu (Kelly) Luo,[1] Kathleen M. McCreary,[2] Simranjeet Singh,[1] Iwan B. Martin[1], Edward J. Cichewicz Jr.[1], Berend T. Jonker,[2] Roland K. Kawakami[1]\**

[1]*Department of Physics, The Ohio State University*, Columbus OH 43210, USA
[2]*Naval Research Laboratory*, Washington DC 20375, USA

\*e-mail: kawakami.15@osu.edu

## I. Additional characterization of the CVD-grown monolayers of WS$_2$

### I.A. N-type doping

The CVD grown material used in this study was determined to be n-type based on transport measurements. Two-probe field effect transistor (FET) devices were fabricated from WS$_2$ monolayers on SiO$_2$/Si(100) substrates. The n-doped Si substrate served as a back gate for electrical measurements. Electrical contacts were defined using electron beam lithography and Au contacts with Ti adhesion layers were deposited via electron-beam evaporation. A constant bias of +12 V was applied between the two metal electrodes while the current was measured as a function of back gate voltage. Measurements were obtained under vacuum (< 10$^{-5}$ Torr) and at 10 K. Figure S1 shows the current as a function of gate voltage, and indicates that the material is n-type. We note the intrinsic conductance (for gate voltage = 0 V) is low, suggesting a low residual n-type doping. The low doping level is also suggested by the small Stokes shift (<2 nm) when comparing the reflectivity spectrum and PL spectrum (see Figure S2), as Stokes shifts are known to increase monotonically with the doping level[1]. The low doping level is also consistent with the PL spectrum, which is dominated by neutral exciton emission ($X^0$), with only minor contributions from charged trion emission (X-).

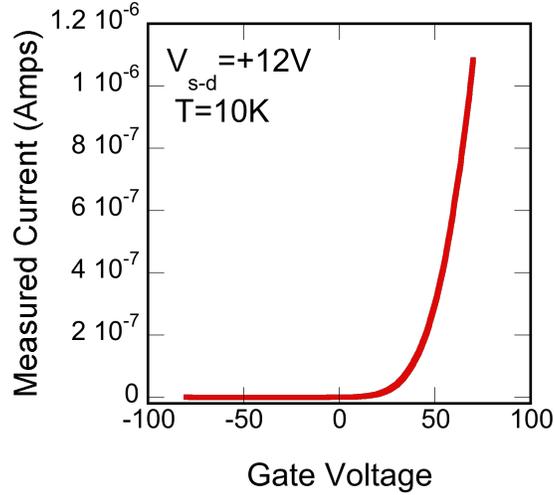

**Figure S1.** Measured current versus gate voltage on a two-probe FET device. Measurements were done at 10 K with a source-drain voltage of +12 V.

## *I.B. PL and reflectivity spectroscopy*

The PL emission energy in our CVD grown samples is considerably different than previously reported on mechanically exfoliated samples. While mechanical exfoliation is typically carried out at room temperature, we perform CVD synthesis at 825°C. As the sample cools to room temperature from the growth temperature, monolayer $WS_2$ and the supporting substrate contract at different rates, imparting strain to the $WS_2$ monolayer, subsequently reducing the band bap and red-shifting the emission energy of the exciton. While we observe a small amount of sample-to-sample variation, room temperature PL emission is near 633 nm (Figure S2), consistent with previously reported room temperature PL energies of 625 nm – 636 nm [2-4] for monolayer $WS_2$ synthesized under conditions similar to our growth parameters (i.e. ambient pressure, $WO_3$ and sulfur precursor, $T_{growth}$> 700°C, $SiO_2$/Si growth substrate).

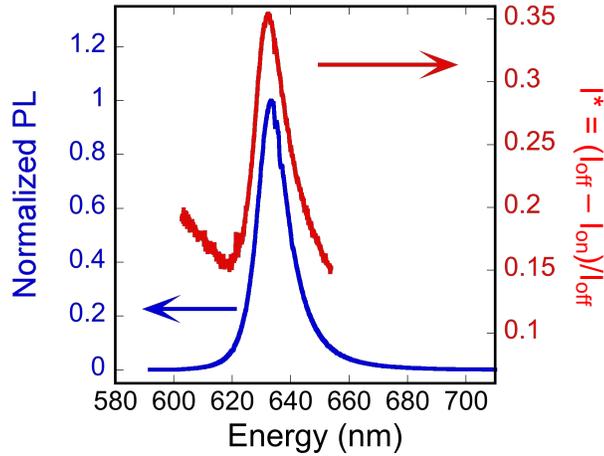

**Figure S2.** PL and reflectivity spectra of the $WS_2$ flakes. These measurements were performed at room temperature.

Figure S2 shows the PL and reflectivity spectra measured at room temperature. To minimize unwanted contributions from the substrate in the reflectivity measurement, the difference between the intensity measured from the bare substrate ($I_{off}$), and from the $WS_2$ sample ($I_{on}$) is obtained then normalized to the substrate intensity, $I^* = (I_{off} - I_{on})/I_{off}$. The peak position measured for PL and reflectivity are nearly coincident, confirming the PL arises from excitonic emission rather than low-energy defect related recombination. The small Stokes shift (<2 nm) is comparable to [5] or better than [6] reported values on mechanically exfoliated $WS_2$.

## II. Wavelength Dependence of TRKR signal

We investigated the wavelength dependence of the TRKR signal and compared it with the wavelength dependence of the PL and reflectance. Supplementary figure S3 displays the PL, differential reflectance, and TRKR signals as a function of wavelength for flake 1 in the main text. The differential reflectance data was taken at 8 K, as was the PL and TRKR (see methods in the main text). The white light was focused down to ~5 μm spot, and the difference between the intensity measured from the bare substrate and the intensity from the sample was normalized to the substrate intensity to minimize substrate contributions, as done for the room temperature data in Supplementary section I.B. The reflectance peak is observed at the same energy as the exciton emission and the TRKR signal. This is expected, as the exciton should exhibit absorption, and the presence of a Kerr rotation in a semiconductor relies on the absorption of circularly polarized light. Neither absorption nor Kerr rotation is observed at the trion (~630 nm) for flakes 1-4 or defect peak (~640 nm) for flakes 1-5. Flake 5, however, does exhibit a Kerr rotation at the trion energy, ~625 nm. Existence of a Kerr rotation signal indicates that the trion absorbs light, which seems to occur when the trion luminescence is large relative to that of the exciton[7]. Photoluminescence from flake 5, main text Figure 1c, shows a very large trion peak, ~20x larger than the neutral exciton (this varies with position).

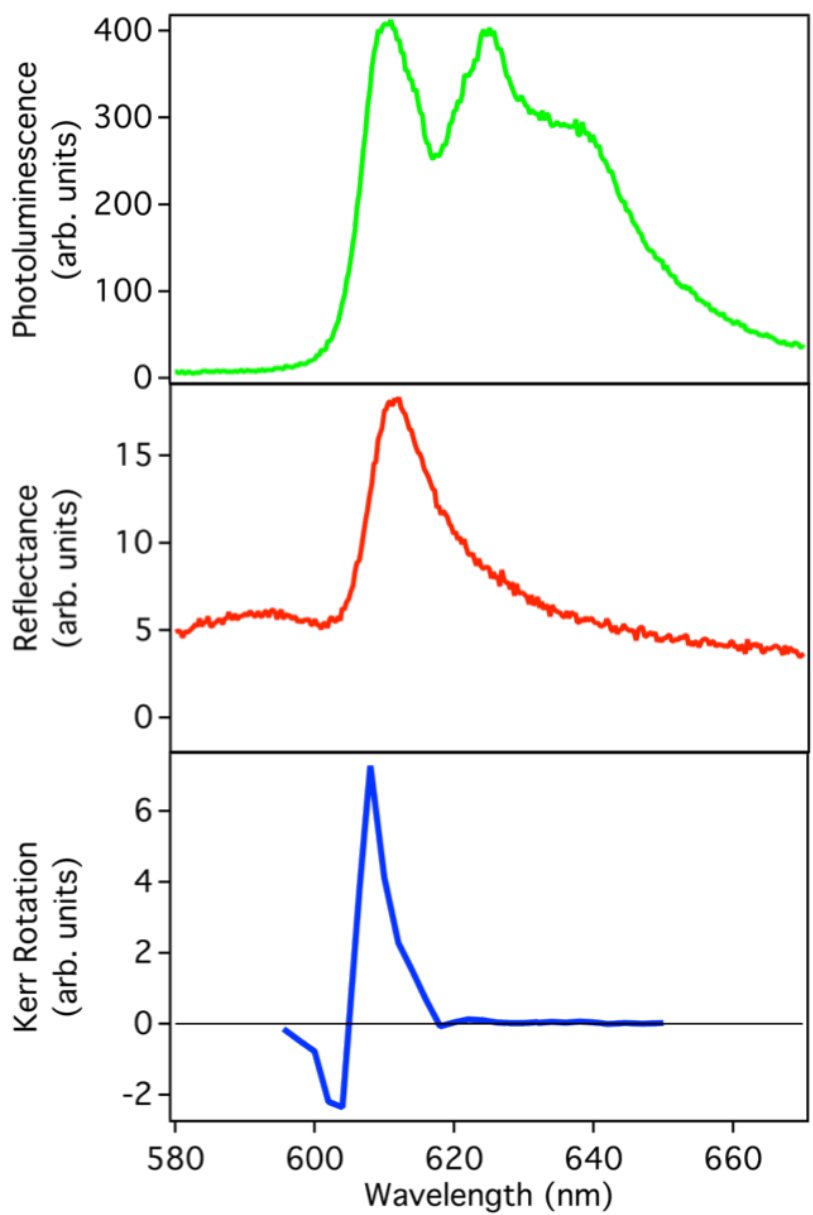

**Figure S3.** Wavelength dependence of the photoluminescence, reflectance, and Kerr rotation, from top to bottom. The Kerr rotation data shows the amount of Kerr rotation present at 100 ps. As can be seen from the three panels, the Kerr rotation only occurs at the energy of the exciton, where there is also a peak in the reflectance, however no Kerr rotation exists at the defect peak, which does not have a corresponding peak in the reflectance spectrum.

### III. Laser Induced Spot on Flake 1

Throughout this study, we observed differing levels of laser induced changes to $WS_2$ flakes. Other studies have also noted that $WS_2$ samples can change over time, both with and without laser exposure [8,9]. Over time, all flakes show some level of aging, but some flakes were affected much faster than others. Flake 1 was especially sensitive to laser exposure. The anticorrelated spot, circled in white in the main text Figure 3a, developed after a relatively short period of exposure to the ultrafast laser. We investigated this specific spot for ~45 min to perform time delay scans for characterizing the spin/valley lifetime. Afterwards, the spot showed a stronger TRKR signal, with suppressed PL. We do not fully understand the mechanism in which the laser exposure was able to induce an anticorrelation, however we suspect that cleaning the surface due to laser annealing or photo-doping may be contributing.

### IV. Linecuts of TRKR and PL on Flake 5

Supplementary Figure 4 shows a series of TRKR delay scans and PL spectra taken along a linecut within the triangular island (flake 5) shown in the main text Figure 2, as indicated in Supplementary Figure 4a. The linecuts elucidate a very important trend, showing that in regions with the strongest PL, the TRKR has a fast decay (<10 ps) and shows very little amplitude in the long-lived spin state. On the other hand, regions with high spin density in the long-lived state have much weaker PL.

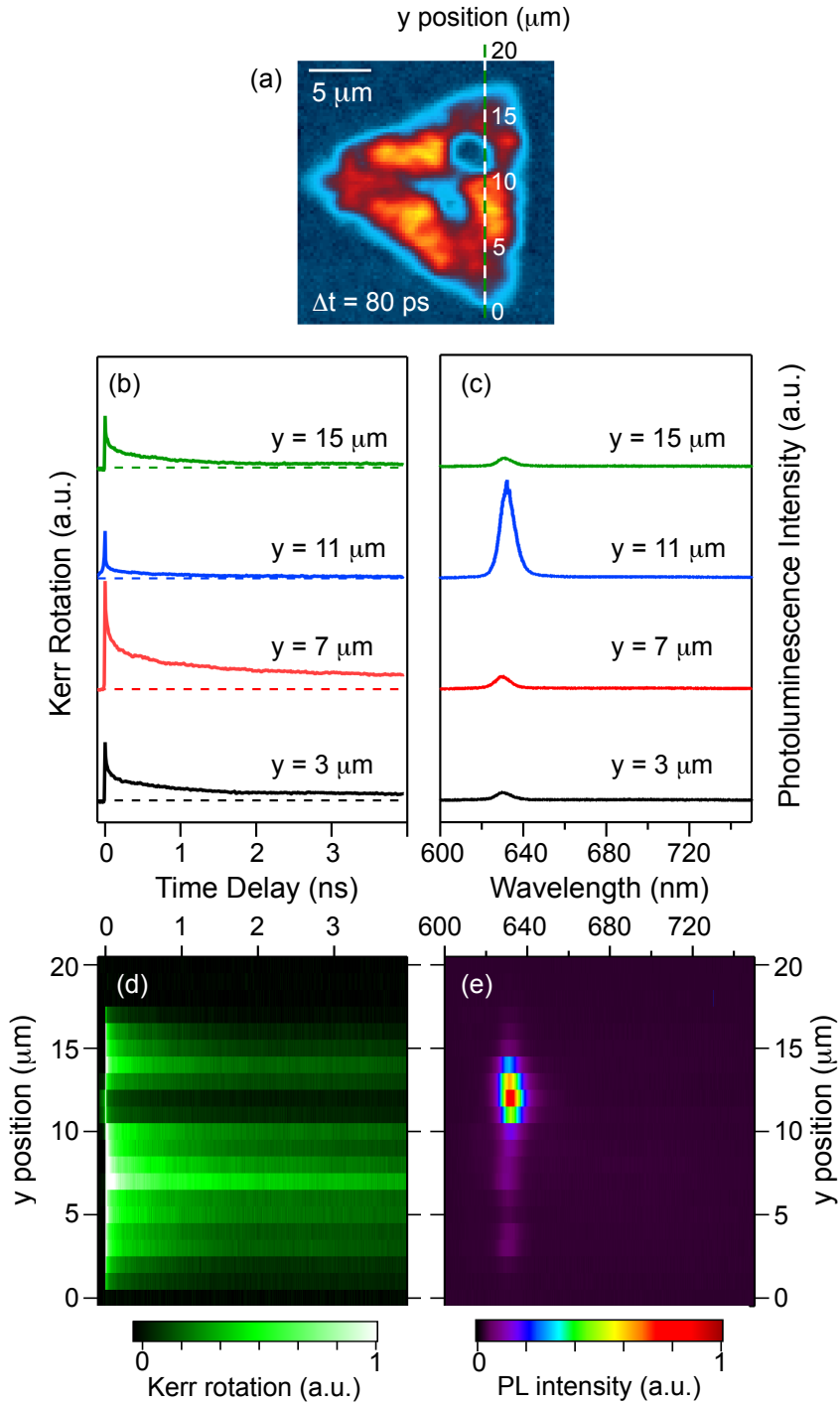

**Figure S4. Anticorrelation of photoluminescence and time resolved Kerr rotation. a,** The dashed line indicates the line-cut where TRKR and PL are compared. **b-c,** TRKR delay scans and PL spectra measured at 6 K at representative points along the line-cut. The position with the brightest photoluminescence has lowest spin density. **d-e,** Detailed spatial dependence of TRKR and PL along the line-cut.

## VI. Temperature dependence of TRKR signal in $WS_2$

The temperature dependence of the TRKR signal in $WS_2$ is shown in Supplementary Figure S5. The TRKR signal was measured as described in the main text, and the long spin/valley lifetime, as given by an exponential fit, was plotted as a function of temperature. The TRKR signal is robust until ~130 K, at which point it drops to below 200 ps. The inset of Supplementary Figure S5 shows representative scans at different temperatures.

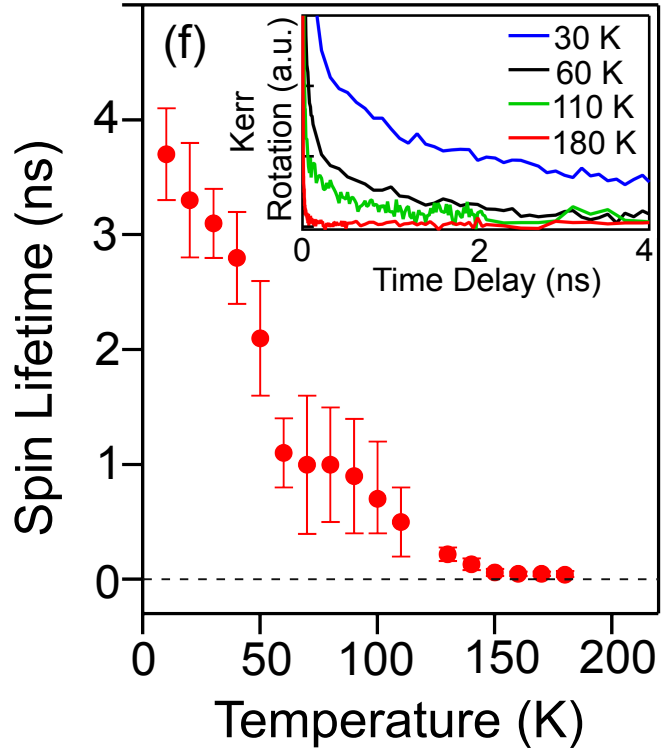

**Figure S5.** Spin lifetime as a function of temperature. The inset shows representative delay scans at different temperatures.